\documentclass{article}
\pdfoutput=1

\usepackage{arxiv}

\usepackage[utf8]{inputenc} 
\usepackage[T1]{fontenc}    
\usepackage{hyperref}       
\usepackage{url}            
\usepackage{booktabs}       
\usepackage{amsfonts}       
\usepackage{nicefrac}       
\usepackage{microtype}      
\usepackage{graphicx}
\usepackage[caption=false]{subfig}
\usepackage{doi}
\usepackage{multirow}
\usepackage{multicol}

\title{MetaEnhance: Metadata Quality Improvement for Electronic Theses and Dissertations of University Libraries}

\author{Muntabir Hasan Choudhury\\
	Old Dominion University\\
	Norfolk, VA 23529 \\
	\texttt{mchou001@odu.edu} \\
    \And
    Lamia Salsabil\\
    Old Dominion University\\
    Norfolk, VA 23529\\
    \texttt{lsals002@odu.edu} \\
	\And
    Himarsha R. Jayanetti\\
    Old Dominion University\\
	Norfolk, VA 23529 \\
	\texttt{hjaya002@odu.edu} \\
	\And
    Jian Wu\\
	Old Dominion University\\
	Norfolk, VA 23529 \\
	\texttt{j1wu@odu.edu} \\
	\And
	William A. Ingram \\
	Virginia Polytechnic Institute and State University \\
	Blacksburg, VA 24061 \\
	\texttt{waingram@vt.edu} \\
	\And
	Edward A. Fox \\
	Virginia Polytechnic Institute and State University \\
	Blacksburg, VA 24061 \\
	\texttt{fox@vt.edu} \\
}

\renewcommand{\shorttitle}{MetaEnhance: Metadata Quality Improvement for ETDs of University Libraries}

\begin{document}
\maketitle

\begin{abstract}
Metadata quality is crucial for digital objects to be discovered through digital library interfaces. However, due to various reasons, the metadata of digital objects often exhibits incomplete, inconsistent, and incorrect values. We investigate methods to automatically detect, correct, and canonicalize scholarly metadata, using seven key fields of electronic theses and dissertations (ETDs) as a case study. We propose MetaEnhance, a framework that utilizes state-of-the-art artificial intelligence methods to improve the quality of these fields. To evaluate MetaEnhance, we compiled a metadata quality evaluation benchmark containing 500 ETDs, by combining subsets sampled using multiple criteria. We tested MetaEnhance on this benchmark and found that the proposed methods achieved nearly perfect F1-scores in detecting errors, and correctness F1-scores ranging from 0.85 to 1.00 for five of seven fields. The codes and data are publicly available at \url{https://github.com/lamps-lab/ETDMiner/tree/master/metadata_correction}.
\end{abstract}

\keywords{Digital Libraries \and Scholarly Big Data \and ETD \and Metadata Quality}

\section{Introduction}
Metadata represents a key aspect of digital objects. Improving metadata quality for digital library (DL) objects is a long-standing problem. Although DL systems have adopted Dublin Core (DC) to standardize metadata formats (e.g., ETD-MS v1.1), Bui and Park et al. \cite{Bui_Park_2013} provided an analysis of 659 metadata item records that showed frequent inaccurate, incomplete, and inconsistent metadata element usage. To address metadata quality issues, in one survey paper \cite{TANI20131194}, the authors discussed the overlaps of quality assessment frameworks defined metadata quality parameters, dimensions, and metrics. One of the existing methods to improve DL metadata is crowd-sourcing, letting users correct metadata errors \cite{wu2014collaboratecom}. But this method has two drawbacks -- a) it is difficult to control the user population, and b) it is neither fast nor scalable. While most existing frameworks explored methods relying on semi-automatic approaches or manual correction, with the advancement of AI, it is possible to explore natural language processing (NLP) and computer vision (CV) methods to improve metadata quality by automatically detecting, correcting, and canonicalizing metadata. Due to the heterogeneous nature of digital objects, it is challenging to design a single system that fixes all metadata fields. Accordingly, in a pilot study, we use metadata of Electronic Theses and Dissertations (ETDs).

ETDs are scholarly documents that represent students' research and demonstrate their ability to independently conduct and communicate research findings and meet the requirements for an academic degree. ETDs are usually hosted by university libraries or centralized online repositories such as  ProQuest. The metadata of ETDs were originally input into the system by students, faculty, or library staff. So presumably, they should be complete, consistent, and accurate. However, upon inspecting metadata downloaded from several university libraries, we noticed many ETD repositories are accompanied by incomplete, inconsistent, or even incorrect metadata. Findings reported in the paper \cite{sami-etd_crawling}, at least 43\% of department fields and 12\% of year fields were empty. Low metadata quality may significantly harm the discoverability of digital libraries.

In this paper, we propose a framework called MetaEnhance, aiming at improving ETD metadata quality by automatically filling in missing values, detecting and correcting errors, and canonicalizing surface names. We quantitatively demonstrated the effectiveness of our system in improving the seven key metadata fields, including title, author, university, year, degree, advisor, and department -- which are ubiquitous in ETDs. Our contributions are the following. (a) We proposed MetaEnhance to improve the metadata of ETDs using AI methods. (b) We created a new benchmark using real-world ETD metadata to evaluate metadata quality improvement methods. (c) Our proposed framework achieved a remarkable performance to improve metadata quality in the benchmark data. The code and data are available at the anonymous GitHub\footnote{\url{https://github.com/lamps-lab/ETDMiner/tree/master/metadata_correction}}.

\section{Related Work}
Several digital libraries allow users to manually correct metadata. For example, when Microsoft Academic was online, it allowed users to change header information, including titles, authors, year, DOI, conference, journal, URL, and abstract. Wu et al. proposed user corrections as a form of crowd collaboration, providing an efficient way to improve metadata quality for CiteSeerX \cite{wu2014collaboratecom}. The authors inspected the correction history and showed that user correction was a reliable source of high-quality metadata. For the author field, it is 100\% correct and for the title field, it is 94.55\%. However, the paper only examined two metadata fields.

Park et al. \cite{park09} argued the existence of inconsistent metadata because many different data providers may not strictly follow the DC schema. The author discussed and compared several methods to measure metadata quality and emphasized the most commonly used criteria, including accuracy, completeness, and consistency. The author compared published methods that proposed guidelines, best practices, and approaches for quality assurance. The author advocated the development of a framework for assessing quality, and mechanisms to improve metadata quality. However, to our best knowledge, no AI-based  frameworks have been proposed and implemented to improve metadata quality for ETDs.

Our work is different from existing work because we focus on using AI methods to \emph{automatically} improve metadata quality, which is more scalable than manual approaches.

\begin{figure}[t]
    \centering
   \includegraphics[width=0.47\textwidth]{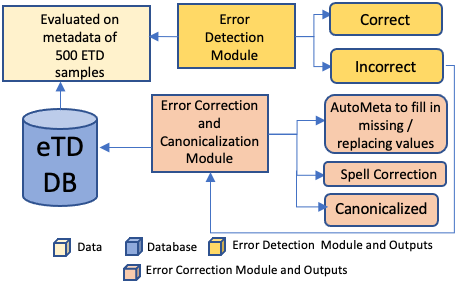}
   \caption{MetaEnhance Framework.}
   \label{fig:framework}
\end{figure}

\begin{table}
    \centering
    \caption{Distribution of ETD errors in the dataset used in this paper.
    The \#Canonical column shows the number of values that could be canonicalized for each field. The canonicalization count includes the values after missing fields are inserted and errors are corrected. The title and author fields do not contain surface names that should be canonicalized.}
    \begin{tabular}{c|c|c|c|c}
    \toprule
    {\bf Field} & {\#Missing} & {\#Canonical} & {\#Spell} & {\#Incorrect}\\
    \midrule
    Title & 0 & 0 &  0 &1 \\
    Author & 2 & 0 & 0 & 0\\
    Advisor & 150 & 35 & 0 & 0  \\
    University & 6 & 43 & 0 & 0\\
    Year & 172 & 1 & 0 & 0\\
    Degree & 156 & 82 & 0 & 4 \\
    Department & 269 & 85 & 2 & 0\\
    \bottomrule
    \end{tabular}
    \label{tab: etddistribution}
\end{table}


\section{Methodology}
Figure~\ref{fig:framework} illustrates the framework, which is comprised of three main modules: error detection, correction, and canonicalization. We describe each module in the following sections.

\subsection{Error Detection}
\label{errorDetectionSection}
Metadata usually has different data types and meanings. We detect three types of errors -- missing values, incorrect values 
and misspellings. If missing values are detected, the module switches directly to the correction module, in which the metadata is parsed from the ETD document using a machine learning method that was previously developed and validated. We then inserted the metadata into the corresponding field. If the field is not empty, the module checks whether the content contains errors. Below, we elaborate on the error detection methods. Each field in our study corresponds to the field in the DC \footnote{\url{https://ndltd.org/wp-content/uploads/2021/04/etd-ms-v1.1.html}} (in parentheses after field names). 

\paragraph{{\bf Title} (dc.title)} This field might contain incorrect ETD titles, such as, ``DMA Recitals''. We adopted the classifier proposed by Rohatgi et al. \cite{rohatgi2021jcdl} to automatically detect incorrect titles. Each string in the title field is represented by the following features: the number of tokens, the number of special characters, the number of capital letters, consecutive punctuation, stop words, the minimum, maximum, and median TF-IDF. The classifier was evaluated on the SciDocs \cite{specter2020cohan} dataset and achieved an F1 score of 0.96.
\paragraph{{\bf Author and Advisor} (dc.creator and dc.contributor)} 
We adopted a named entity recognition model implemented in the FlairNLP \cite{flair-2018} package to detect incorrect author and advisor names automatically. This model was pre-trained and evaluated on the \emph{Ontonotes}\footnote{\url{https://catalog.ldc.upenn.edu/LDC2013T19}} dataset and achieved an F1 of 0.90. An error is detected if the author/advisor name or any part of it is classified as a type other than PERSON.
\paragraph{{\bf University } (thesis.degree.generator)} The university field contains different surface names referring to the same university. For example, ``Johns Hopkins University" is abbreviated as ``jhu" or ``JHU". We built a dictionary\footnote{\label{note4}\url{https://en.wikipedia.org/wiki/List_of_colloquial_names_for_universities_and_colleges_in_the_United_States}} which consists of 832 names of universities of the United States and their acronyms. We checked the surface values against this dictionary and determined if any value was incorrect. Note that the field was marked as an error so the correction module could inspect it and decide whether it should be corrected or canonicalized.
\paragraph{{\bf Degree} (thesis.degree.name)} For the degree field, our database record shows both acronyms and incorrect metadata. For example, the word ``history'' may appear in this field. We used a dictionary based method by compiling a dictionary containing 234 degree names and their acronyms based on the degree naming convention of 
\emph{DegreeAbbreviations}\footnote{\label{note1}\url{https://abbreviations.yourdictionary.com/articles/degree-abbreviations.html}}\footnote{\label{note2}\url{https://degree.studentnews.eu/}}. Any value for a degree not found in this dictionary was identified as an error. Note that these degree names may not be ``wrong''. They are marked as ``errors'' so further modules will inspect whether they are errors or need to be canonicalized.
\paragraph{{\bf Department} (thesis.degree.discipline)} We observed numerous misspellings for department fields such as ``College of Muisc'', ``scool of Music'', ``Graduhte Studies in English''. To detect spelling errors in surface names, we used the Python library \emph{pyspellchecker}\footnote{\label{note6}\url{https://pypi.org/project/pyspellchecker/}}. It uses Levenshtein distance to detect spelling errors. If the editing distance between the original word and the field value is 2, it is considered a spelling error.

\paragraph{{\bf Year} (dc.date.issued)} The format of this field is inconsistent across libraries, such as:``mm-dd-yyyy'' or ``yyyy-mm-dd''. We used the Pandas built-in parser from ``dateutil.parser.parse'' \footnote{\url{https://dateutil.readthedocs.io/en/stable/parser.html}}, which can perform generic parsing of dates in almost any string. We verified if the specific date field was valid using this parser, and then checked against a dictionary listing the year range from 1880 to 2023 to detect errors. If we marked the value as an error, the correction module could inspect it and decide whether it should be corrected or canonicalized.

\subsection{Error Correction \& Canonicalization (ECC)}
If an error was detected in the degree field, we attempt to canonicalize the value by searching in the DegreeAbbreviations dictionary. The ECC module contains two types of corrections, depending on the errors detected in the previous model, in addition to canonicalizing entity surface names. These corrections are: (a) filling in missing values using AutoMeta \cite{choudhury2021automatic} and (b) correcting misspellings and incorrect values.

One challenge of the data correction task is the information source that can be used for filling in missing values and overwriting error values. The MetaEnhance system integrates an existing framework called {\bf AutoMeta} \cite{choudhury2021automatic}, a software framework that automatically extracts seven key metadata fields by combining visual and text features from ETD cover pages using conditional random fields (CRFs). The model was tested on a corpus of 500 ETDs (different from this paper) and achieved 81.3\%--96\% F1 scores depending on the field. We used the best model to extract metadata from ETDs used in this paper and fill in missing values for seven metadata fields.
{\bf Canonicalization} involves converting data with multiple possible surface names into a ``standard" form. We canonicalized the values of the advisor, university, department, degree, and year fields detected by the previous model and the AutoMeta \cite{choudhury2021automatic} result.

\paragraph{{\bf Title}}
If any incorrect title is detected, we overwrite it with the title extracted by the AutoMeta \cite{choudhury2021automatic} for that specific ETD.

\paragraph{{\bf Author and Advisor}} 
If an author or an advisor name was detected as an error, we overwrite the field using corresponding fields extracted using AutoMeta \cite{choudhury2021automatic}. We also observed that advisor names, such as, ``Mark Pankow, Co-Chair'' or ``Andrew Mathew Jr., Committee Member'' needs to be parsed. According to DC, Co-Chair is a role (dc.contributor.role) of a member in the thesis committee. We used regular expressions to parse the surface value and then stored the value in a separate column.

\paragraph{{\bf University}}
If the university name was detected as an error, we employed a dictionary-based method by matching a university name against the university dictionary. We normalized both the field name and the colloquial names by converting all letters to upper case and stripping off punctuation marks, and then searching for the surface name in the dictionary to see if it is colloquial. The full name was then used to replace the surface name in this field. Table \ref{tab:acronym} shows examples of colloquial and official names.  If any incorrect university is detected, we overwrite the incorrect university with the title extracted by AutoMeta \cite{choudhury2021automatic}.

\begin{table}
\centering\small
\caption{Examples of acronym/colloquial names and corresponding expanded names for university, degree, and department fields. \label{tab:acronym}}
\begin{tabular}{c|c|p{3cm}}
\toprule
 {\bf Field} & {\bf Acronym/Colloquial}& {\bf Full/Expanded Name}  \\
 \midrule
  University   & GEORGIA TECH, GT, GIT & Georgia Institute of Technology\\
  \midrule
  Degree & MPHIL, M PHIL, PHM & Master of Philosophy\\
  \midrule
  Department & MSE, MSCE & Materials Science and Engineering\\
  \bottomrule
\end{tabular}
\end{table}

\paragraph{{\bf Degree}} 
If an error was detected in the degree field, we attempt to canonicalize the values by searching it against the \emph{DegreeAbbreviations} dictionary. We first normalized field values that involved converting degree metadata to uppercase letters and removing all punctuation marks. We then replaced acronyms with their expanded forms. If any incorrect degree is detected, we overwrite the incorrect degree with the degree extracted by AutoMeta \cite{choudhury2021automatic}.

\paragraph{{\bf Department}} 
Department names can have different forms. For example, ``Department of Computer Science", ``Dept of CS'', and "CS Department" all map to the same entity. For this field, we first correct spelling errors and then disambiguate department names and canonicalize them into full names.

The Python library \emph{pyspellchecker}\textsuperscript{\ref{note6}} was used to identify and correct {\bf spelling errors}. The library captures errors using Levenshtein distance (see
Section \ref{errorDetectionSection}). All permutations (e.g., insertions, deletions, replacements, and transpositions) are compared to known words in a word frequency list. Words that appear more often are considered correct spelling.

To {\bf canonicalize department names}, we compiled a comprehensive list of 232 different academic department names and their acronyms using the official \emph{Abbreviations and Symbols from Boston University}\footnote{\label{note3}\url{https://www.bu.edu/academics/bulletin/abbreviations-and-symbols/}}. We normalized all the surface names and the acronyms on the list. These surface names were compared with the acronyms on the list to see if they were abbreviated forms. If a match was found, the matching surface name was replaced with its corresponding full name from the list. To disambiguate such surface names, we developed a model using SentenceTransformers \cite{reimers2019sentence}, which generates embeddings of surface names and a department full names with SentenceTransformer and then measures cosine similarity of the surface names against the dictionary set. We observed that 91\% of the records with $cosine-similarity \geq 0.90$ provide correct matches. Table~\ref{tab:acronym} shows an example that our model identifies the similarity and maps them to corresponding full names.

\paragraph{{\bf Year}}
If an error was found in the year field, we
used the value from
AutoMeta \cite{choudhury2021automatic}. To canonicalize the ``year'' surface values, we used regular expressions to identify inconsistent date formats. Then we utilized the Pandas built-in parser from ``dateutil.parser.parse'', which outputs three date fields, including ``year'', ``month'', and ``date''. We stored the values in three separate columns.

\subsection{Database Update \& Version Control}
We replaced the erroneous values (e.g., missing values, colloquial values, misspellings, and incorrect values) of the metadata with corrected values in the database. The detection and correction modules rely on the AI algorithm, which may not be 100\% accurate. It is necessary to add a version control module to keep track of changes and roll back in case a record was later found corrected by mistake. Because the revision of metadata could happen on any field, one question in designing such a module is whether to back up the entire record or only the revised field.  We decided to backup the entire record if any field gets revised for the following reasons. Although backing up a field saved more space and allowed us to keep track of the exact metadata that is revised, the database design is more complicated, and rolling back to the previous values is error-prone. Backing up the entire record is more intuitive, and it is straightforward to identify which field is revised by comparing different versions of records. Each revision is attached with a version number and a timestamp. The version control is implemented by MySQL but can be easily transferred into other database systems.

\section{Evaluation and Results}
\subsection{Dataset}
To evaluate MetaEnhance, we compiled a corpus containing metadata from 500 ETDs selected from 533,047 ETDs by crawling 114 US university libraries, including full text in PDF format, and metadata from university library repositories and ProQuest. Data collection was based on a software framework, which harvested ETDs and their metadata via the Open Archives Initiative protocol (OAI-PMH) or sitemaps \cite{sami-etd_crawling}. To mitigate selection bias against samples that are under-represented in certain dimensions (e.g., a random sampling would be biased against ETDs of minority universities), we select  4 ETD subsets based on 4 different criteria, and then combine them. Because the purpose of this data is to evaluate the metadata quality improvement system, for each criterion, except for the title and author fields, we selected ETDs with missing values for the remaining fields. The selection criteria ensure that we cover less biased samples with diverse properties. Table~\ref{tab: etddistribution} illustrates the distributions of ETDs in different feature spaces.
\begin{itemize}
    \item {\bf Random:} We randomly sampled 100 ETDs from our collection.
    \item {\bf University:} We first randomly selected 10 universities and then 10 random ETDs from each university. 
    \item {\bf Year:} We randomly sampled 100 ETDs each year in 2010--2019. 
    \item {\bf Department:} We randomly sampled 6 STEM and 4 non-STEM disciplines. Then we randomly sampled 10 ETDs of each category. 
    \item {\bf Degree:} We randomly sampled five-degree names and then randomly selected 20 ETDs from each degree. 
\end{itemize}

\begin{table}
    \centering\small 
    \caption{Performance of Error Detection (ED) and Error Correction and Canonicalization (ECC).}
    \begin{tabular}{c|c|c|c|c|c|c}
    \toprule
    \textbf{Field} & {\bf $\mathbf{P}_{\rm ED}$} & {\bf $\mathbf{R}_{\rm ED}$} & {\bf $\mathbf{F1}_{\rm ED}$} & {\bf $\mathbf{P}_{\rm ECC}$} & {\bf $\mathbf{R}_{\rm ECC}$} & {\bf $\mathbf{F1}_{\rm ECC}$}\\
     \midrule
     {Title} & {0.997} & {1.0} & {0.998} & {0.0} & {0.0} & {0.0} \\
     \midrule
      {Author} & {0.996} & {1.0} & {0.997} & {0.0} & {0.0} & {0.0}\\
     \midrule
      {Degree} & {1.0} & {1.0} & {1.0} & {0.980} & {1.0} & {0.980}\\
     \midrule
      {Department} & {0.996} & {1.0} & {0.997} & {0.970} & {1.0} & {0.980}\\
     \midrule
      {University} & {1.0} & {1.0} & {1.0} & {0.740} & {1.0} & {0.850}\\
     \midrule
      {Year} & {1.0} & {1.0} & {1.0} & {1.0} & {1.0} & {1.0}\\
     \midrule
     {Advisor} & {0.920} & {0.990} & {0.950} & {1.0} & {1.0} & {1.0}\\
     \bottomrule
    \end{tabular}
    \label{tab:errordetection}
\end{table}

\subsection{Error Detection Evaluation}
Errors include missing values, misspellings, and incorrect values (e.g., titles). Table~\ref{tab: etddistribution} shows the number of missing values for each field. Our error detection module correctly detected all the missing values for each field. Depending on the metadata types, the module detects errors differently.
We checked against the ground truth for each field and reported precision, recall, and F1 scores. Table~\ref{tab:errordetection} shows that the university, year, and degree fields achieved perfect recall and precision. Specifically, all 4 incorrect surface values in the degree field were detected. The error detection module achieves $F1>0.99$ for the title, author, and department fields. 

We found one false positive (FP) and one false negative (FN) for the title field. For example, ``DMA Recitals'' was misclassified as a valid title. We also found 2 FPs for the author field. The classifier misclassified the name ``Richmond Orien Manu Wright''. Here, the first name ``Richmond'' was identified as a geographical entity (GPE), while the rest was identified as a PERSON. For the advisor field, our method achieved F1=0.95 with 37 FPs. Specifically, the method misclassified the name ``Mark Pankow, Co-Chair''. While the department classifier correctly detected 2 misspellings (e.g., ``scool'' in the department field), it misclassified 2 surface values. For example, ``Public Health (PMH)'' was classified as an incorrect department name.


\subsection{ECC Evaluation}
The performance of the ECC module relies on the output of AutoMeta \cite{choudhury2021automatic}. 
Although AutoMeta achieved F1-scores of 0.67 -- 0.91 for most fields, it failed to extract advisor field values from most of the ETDs, because most advisor fields do not appear on the cover pages of the ETDs in our corpus.
Further, AutoMeta often erred when applied to ETD cover pages that resulted from low resolution scans.

The title field does not contain missing or incorrect values  (Table~\ref{tab: etddistribution}). The error detection module only detected two missing values for the author field. However, AutoMeta could not extract authors for these two ETDs to fill in those missing values, which leads to zero precision, recall, and F1. For the other fields, the ECC module successfully corrected all missing values. Moreover, the module successfully canonicalized the surface names in degree (e.g., ``Ph.D.''), department (e.g., ``CS''), and university (``JHU'') fields. Table~\ref{tab: etddistribution} shows the total numbers of values needing to be canonicalized. Note that authors and titles do not need to be canonicalized. We canonicalized 
7\%, 6.4\%, 0.2\%, 16.2\%, 16.6\% for the advisor, university, year, degree, and department fields, respectively. Table~\ref{tab:errordetection} shows that we successfully canonicalized all values in year and advisor fields and a high percentage of values in other fields.
When the department and degree fields were incorrectly extracted by AutoMeta \cite{choudhury2021automatic}, they were not canonicalized.
Furthermore, the ECC module successfully corrected 4 incorrect values, and 2 misspellings, for the degree and department fields, respectively.




\section{Conclusion and Discussion}
We developed MetaEnhance, a system to automatically improve ETD metadata using AI methods. We then applied it to our benchmark dataset and quantitatively demonstrated the effectiveness of our system in improving the metadata quality. Overall F1 scores, depending on the metadata fields: MetaEnhance achieved 95\%--99\% in detecting errors, and
corrected 85\%--98\% of errors, e.g., filling missing values, and canonicalizing surface names. One of the limitations of the error detection module is that it may misclassify a valid university or degree if it is not found in the dictionary. In addition, the department field canonicalization only maps the acronyms to the full names when they are included in the Boston University dictionary list of department names. It is possible that universities may not follow exactly the same convention for certain acronyms. The benchmark data can be made more challenging by incorporating more spelling errors and incorrect values, which can be achieved by introducing random noise to true field values.

\section*{Acknowledgement}
Support was provided by the Institute of Museum and Library Services through grant LG-37-19-0078-198.

\bibliographystyle{unsrt}
\bibliography{references}  






\end{document}